\documentclass[preprint,showpacs,preprintnumbers,amsmath,amssymb,floatfix,superscriptaddress,boldsymbol]{revtex4}

\def\b{\mathbf}

\usepackage{amssymb,subeqnarray}
\usepackage{amsmath}
\usepackage{graphicx}
\usepackage{dcolumn}
\usepackage{bm}
\usepackage{color}

\begin{document}

\title{Fluid transport by active elastic membranes}

\author{Arthur A. Evans}
\affiliation{Department of Physics,}
\author{Eric Lauga}
\affiliation{Department of Mechanical and Aerospace Engineering, \\
University of California San Diego,  9500 Gilman Drive, La Jolla, CA 92093, USA.}
\date{\today}

\date{\today}

\begin{abstract}
A flexible membrane deforming its shape in time can self-propel in a viscous fluid. Alternatively, if the membrane is anchored, its deformation will lead to fluid transport. 
Past work in this area focused on situations where the deformation kinematics of the membrane were prescribed. Here we consider  models where the deformation of the membrane is not prescribed, but instead the membrane is internally forced. Both the time-varying membrane shape, and the resulting fluid motion, result then from a balance between prescribed internal active stresses, internal passive resistance, and external viscous stresses.  We introduce two specific models for such active internal forcing: one where a distribution of active bending moments is prescribed, and one where active inclusions exert normal stresses on the membrane by pumping fluid through it.  In each case, we  asymptotically calculate the membrane shape and the fluid transport velocities for small forcing amplitudes, and recover our results using scaling analysis. 

\end{abstract}




\maketitle

\section{Introduction}

Active materials, ranging from living fluids to lipid membranes interspersed with force generating molecular machines, present interesting challenges for modern soft matter physicists \cite{ActiveReview}.  Understanding the dynamics  of materials whose characteristics and responses depend on dynamically varying internal stresses  holds promise for revealing meaningful features of the cellular world. Mechanical feedback between the environment that a cell is immersed in and the fluctuating inner behavior of its internal constituents plays an important role in motility, morphology, and reproduction \cite{Karp99,McMahon:2005p201,Bottino:2002p649,Grimm:2003p648}.   The study of biological membranes, in particular, and the role of morphology in ultimately determining the functionality of a cell, has long generated interested in the scientific community.  The geometry of a cell impacts the proteins embedded in its surface \cite{Huang:2006p115}, and the shape fluctuations of an active membrane yields insight about the activity within \cite{Lee:1999p477,BrochardLennon1975,Manneville,Veksler:2007p136}.

In recent decades cell locomotion has occupied a great deal of attention  \cite{Purcell:1977p29,Lauga:2009p421,braybook,bergbook}. 
One of the possible justifications for this interest stems from the fact that self-propelled organisms represent one of the ways in which soft active transport is accessible to our intuition. In all of these cases, and in many others, shape matters.  The deformation of a biological membrane, and the rate at which it occurs, inevitably determines the effect that the internal stress state has on the world around it:  Internal activity competes with dissipative forces arising from viscous fluids, frictional substrates, or other external forces and -- in addition to the particular constitutive relationship ruling the behavior of the membrane itself -- the final result is the shape of the body.

Focusing on cellular motility, and swimming in particular, the only external stress is that exerted by the viscous fluid on the deforming surface.  Provided that the deformation of the membrane is not time-reversible, the body performs work against the fluid and generates a macroscopic velocity \cite{Purcell:1977p29}.  Dual to this problem is  fluid pumping, wherein an actively deforming tethered membrane transports fluid, rather than propelling itself through the bulk.  This aspect of fluid transport is the focus of the current paper.

To understand the origin of  fluid transport by a beating membrane, one only needs to know the deformation of the surface and the fluid properties; this is, in fact, how previous work on the subject has been developed, either to model actual organisms or to provide concepts for locomotion that do not occur in Nature \cite{taylor51,Lauga:2009p421,LaugaViscoelastic,ss96,Evans2010}.  If the kinematics of a membrane deformation are prescribed,  the transport characteristics require  thus only solving the fluid mechanics problem \cite{taylor51}. 

A more physically-relevant model  would start from knowledge of the internal forcing, and then both the deformation and the transport would be solved for at the same time. Recently there have been attempts to prescribe not merely the kinematics, but instead the internal dynamics of a deforming body as model for the physics of axonemal beating in eukaryotic cells   \cite{Camalet,Fu:2008p518}.  
The physical problem becomes then: given an internal state, and a dynamic evolution equation, what are the macroscopic results? Past work has focused on  active filaments, and our present study  extends thus this dynamical analysis to membranes.

In this manuscript we present a model for  the internal force generation in an active membrane. Introducing two  models for internal actuation, and taking advantage of the asymptotic limit of small forcing, we analytically derive the membrane deformation from its linear response, and then use the deformation to deduce the (quadratic) fluid transport. Our results are recovered by  scaling arguments, which allow us to intuitively quantify how the three-way balance between internal forcing, passive (elastic) constitutive modeling and external viscous forcing impacts fluid transport.

\section{Transport by general deformation of a sheet}
\label{sec:II}
\subsection{Setup}
\begin{figure}[t!]
\includegraphics[width=.5\textwidth]{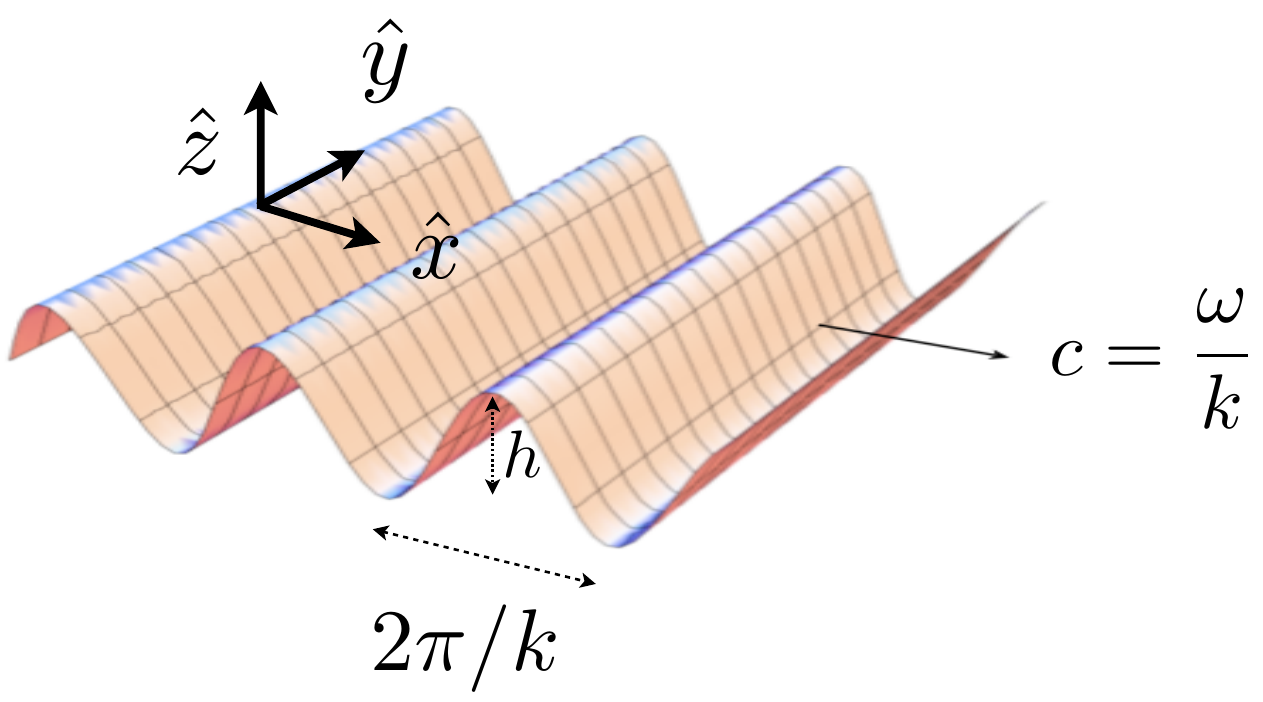}
\caption{\label{schematic} Generalized Taylor swimming sheet passing a traveling wave in the positive $x$ direction with constant wave speed $c=\omega/k$.  The wavelength is $2\pi/k$ and the 
height of the membrane denoted $h(kx-\omega t)$. 
In the  reference frame of the sheet, the material points undergo transverse displacements, while at infinity a uniform pumping flow $U$ develops.}
\end{figure}

For the microscopic regimes that we are interested in the fluid flow is well modeled by the incompressible Stokes equations,  $\nabla\cdot\mathbf{\sigma}=0$, $\nabla\cdot \b{u}=0$, where $\b{u}$ is the fluid velocity, and $\sigma$ is the fluid stress tensor. For this work we consider only Newtonian fluids, such that the first condition becomes $\nabla p=\mu\nabla^2\mathbf{u}$, where p is the pressure and $\mu$ is the shear viscosity.   We consider an infinite, two-dimensional sheet that passes a traveling wave of arbitrary shape $h$ over its surface (see Fig.~\ref{schematic} for notation), in the absence of thermal fluctuations.  If there is no variation in the y-direction then the fluid is two-dimensional and a streamfunction $\psi$ such that $\b{u}=\psi_z\hat{x}-\psi_x\hat{z}$ can be defined. 

For an arbitrarily shaped traveling waveform $h(kx-\omega t)$, we apply a no-slip boundary condition to the sheet to get
\begin{subeqnarray}
 u_x&=&\frac{\partial \psi}{\partial z}|_S=0,\\
 u_z&=&-\frac{\partial\psi}{\partial x}|_S=-\frac{\partial h}{\partial t},
\end{subeqnarray}  
where these conditions must be applied on the material itself, $S$.  This is precisely what leads to geometric nonlinearities and precludes a full analytical solution to  the present problem.  

\subsection{Fluid pumping}
We expand the waveform as  $h=\epsilon h^{(1)}+\epsilon^2h^{(2)}+...$ where $\epsilon$ is a small parameter denoting the magnitude of the wave amplitude.  The stream function $\psi$ is expanded similarly.

To leading order, we write  $h^{(1)}=\Re\{\sum_n{b_ne^{in(kx-\omega t)}}\}$ and, following Childress  \cite{Childress},  solve for the stream function to obtain
\begin{gather}
\psi^{(1)}=\Re\{\sum_n{\frac{\omega}{k}b_n(1+nkz)e^{-nkz}e^{in(kx-\omega t)}}\}.
\end{gather}
At this  order there can thus be no flow far from the sheet: the $h\rightarrow -h$ symmetry demands that any expansion of the velocity $U$ be symmetric in powers of $h$.  

At second order, then, we find that
\begin{gather}
\psi_z^{(2)}(x,0)=-\psi_{zz}^{(1)}(x,0)\Re\left\{\sum_n{inkb_ne^{in(kx-\omega t)}}\right\}.
\end{gather}
Since  the sheet is periodic, averaging this quantity over one period in space yields the flow at infinity, or the macroscopic fluid transport velocity, and we obtain
\begin{gather}
\label{velocity}
U^{(2)}=\frac{1}{2}\sum_n{\omega k |n b_n|^2}.
\end{gather}
Importantly, we see that the knowledge of only the first order height coefficients, $b_n$, leads to the determination of the fluid transport properties at second order.

\subsection{Stress}
In the following section we will invoke local force balance at leading order to determine the membrane shape  and thus we need to know the distribution of stress from the fluid. The pressure at first order is given by 
\begin{gather}
p^{(1)}=-2\mu\omega \Re\{\sum_n{ inkb_ne^{-nkz}e^{in(kx-\omega t)}}\},
\end{gather}
while the components of the fluid stress are
\begin{subeqnarray}
{\sigma}^{(1)}_{zz}&=&-p^{(1)}+2\mu\frac{\partial^2\psi^{(1)}}{\partial x\partial z} ,\\
{\sigma}^{(1)}_{xz}&=&2\mu\left(\frac{\partial^2\psi^{(1)}}{\partial x^2}-\frac{\partial^2\psi^{(1)}}{\partial z^2}\right).
\end{subeqnarray}

\section{Active membrane mechanics}

We now proceed to derive the  dispersion relations for two models of active elastic sheets that will provide a quantitative bridge between the microscopic formulation and the macroscopic flow. 

In general the internal forces (i.e.~the forces not originating with the viscous fluid) will consist of a passive elastic response and an active component.  The general enthalpy functional that describes the internal energetic state of the membrane is given by \cite{Helfrich1973}
\begin{gather}
\label{enthalpy}
G=\int{\frac{\kappa}{2}(C-C_0)^2}dS+\int{\gamma}dS+G_{act}.
\end{gather}
Here $\kappa$ is the bending rigidity of the membrane, $C$ is the mean curvature, $C_0$ is the so-called spontaneous curvature of the membrane, $\gamma$ is the surface tension and $G_{act}$ is the active contribution to the enthalpy, whose form depends on the particular model of activity, and which we give two examples for below. 

Real biological membranes are complex, containing proteins embedded in the surface, several layers of chemical activity, or possibly even an elaborate scaffolding of interlinked polymer networks (relevant, e.g.,  to the cytoskeleton in eukaryotic cells).  For simplicity, we ignore these effects, as well as possible 
viscous dynamics inside the membranes, and focus on bending energetics  \cite{SeifertReview,PowersRMP}.  In addition, although spontaneous curvature can lead to interesting morphological consequences in cells and vesicles (\cite{Miao1994,Kas:1991p647}), we work with $C_0=0$ and only consider local curvature changes from inclusions in the membrane. The form of the active contribution to the enthalpy, $G_{act}$, depends on the particular method of internal forcing \cite{MMMuller}.  Below we consider two models, focusing on internal bending moments and normal forcing to the membrane respectively.

\subsection{Active bending stresses}

\subsubsection{Setup}

\begin{figure}[t!]
\includegraphics[width=.6\textwidth]{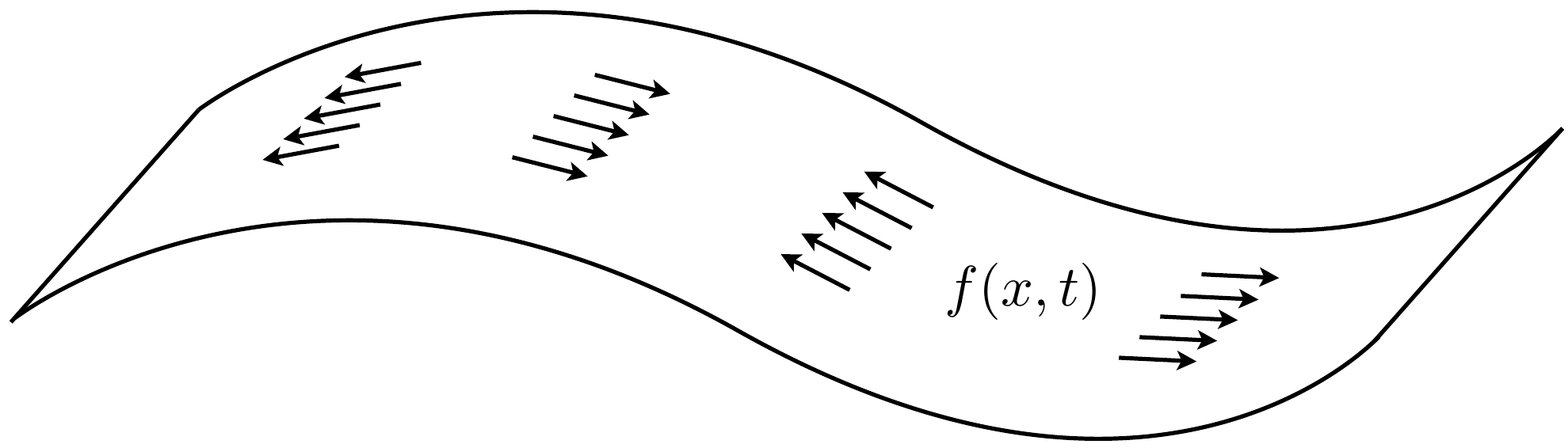}
\caption{\label{tangential} Active membrane where active two-dimensional moments are prescribed with density $f(x,t)$. Normal deformations arise over regions with a gradient in the active stress.}
\end{figure}

In this first model, we assume that there is a distribution of forces acting entirely within the surface of the membrane.  These forces then generate a moment distribution that depends on the thickness of the membrane itself.  We then define an internal, prescribed two-dimensional moment per length (units of force) $f(x,t)$ (see Fig.~\ref{tangential}).  
Balancing this activity with internal passive response and viscous fluid forces yields the instantaneous equations of mechanical equilibrium
\begin{subeqnarray}
\label{forcebalance}
\kappa\nabla^2C +\b{\hat{n}}\cdot\boldsymbol{\sigma} \cdot\b{\hat{n}}|_S&=&\nabla^2 f \,\,\,\,\,\,\, {\rm (normal),}\\
\tau+\b{\hat{t}}\cdot\boldsymbol{\sigma}\cdot\b{\hat{n}}|_S&=&0 \,\,\,\,\,\,\,\, {(\rm tangential)}
,\end{subeqnarray}
where $\tau=\gamma+\kappa C^2$ is the physical tension in the membrane, and  $\b{\hat{t}}$ and $\b{\hat{n}}$ are vector tangent and normal to the membrane respectively.  This  equation is correct for any arbitrary distribution of forces, or any shape of the membrane, as long as $\nabla$ is taken to be the covariant gradient.  For long-wavelength membrane deformation, however, we already solved the fluid mechanics that results in  fluid transport.  In this case the membrane shape can be parameterized by a height field $h(x,t)$, and the curvature $C\approx\nabla^2h$.  To lowest order in the expansion of the height, the equations for the pointwise force balance across the membrane then become
	\begin{subeqnarray}
\label{tangential_EOM}
	\kappa\frac{\partial^4h^{(1)}}{\partial x^4}-\frac{\partial^2 f}{\partial x^2}&=&-p^{(1)}-2\mu\left(\frac{\partial^2\psi^{(1)}}{\partial z\partial x}\right)_S, \\
	\tau^{(1)}&=&\mu\left[\frac{\partial^2 \psi^{(1)}}{\partial z^2}-\frac{\partial^2\psi^{(1)}}{\partial x^2}\right]_S .
\end{subeqnarray}
Using the expression for the first order stream function from the previous section, we find that to first order the tension $\tau^{(1)}=0$: to lowest order in the deformation of the membrane, only normal effects are important \cite{PowersRMP}.

\subsubsection{Scalings}
\label{sec:scaling_tang}

Using scaling arguments  we  derive in this section the expected scaling of the pumping velocity by the active membrane.  In the context of the classical Taylor swimming sheet, the swimming velocity is expected to scale as $U\sim c(bk)^2$, where $c=\omega/k$ is the wave speed. 

Two physical regimes need to be considered, those of ``stiff" and ``floppy" membranes. In the stiff regime, viscous forces are negligible compared to bending resistance, and thus the dynamic balance is between  elastic and active stresses. The elastic stress in a membrane with rigidity $\kappa$,  typical height deformation $b_{{\rm eff}}$, and deformations occurring at typical wavenumbers $k$ scales like $\kappa b_{{\rm eff}}k^4$, while the active stress is on the order of $f_0k^2$.  This yields a value for the effective height of the membrane as $b_{{\rm eff}}\sim f_0/\kappa k^2$. We then expect pumping in the stiff regime, $U_s$, to occur at speed 
$U_s\sim c(b_{{\rm eff}}k)^2\sim\omega f_0^2/\kappa^2 k^3 $. 

In contrast, in the floppy limit the bending resistance is negligible and the dynamic balance is between viscous stresses and internal activity.  The typical shear stress on the sheet scales as $\mu c b_{{\rm eff}}k^2$.  Force balance leads thus to the scaling $f_0 k^2 \sim\mu c b_{{\rm eff}}k^2$, and the deformation is given by $b_{{\rm eff}}\sim f_0  /\mu c $. Fluid pumping in the floppy limit, $U_f$,  is thus predicted to happen with speed $U_f\sim c(b_{{\rm eff}}k)^2\sim f_0^2 k^3/\mu^2 \omega $. Interestingly, in floppy limit, the dependence of the pumping speed on both the sheet frequency and wavenumber is opposite to that in the stiff limit.

To characterize the floppy-to-stiff transition, we introduce the dimensionless group $a=1/k\ell$ where $\ell=(\kappa/\mu\omega)^{1/3}$ is the elasto-viscous penetration length that determines how strongly the membrane shape is effected by the bending resistance versus the viscous forces (similar to the so-called ``Sperm number" used to model viscous locomotion of flagellated organisms \cite{LaugaFlop,Fu:2008p518}). When $a \ll 1$ the membrane is stiff and hence it is energetically prohibitive to introduce an excitation of linear dimension the order of $1/k$, so the viscous forces do not modify the shape of the membrane and  the waveform is a result of the balance between activity and rigidity alone.  In contrast, when $a\gg 1$, the membrane is floppy, and the fluid forces  dynamically balance the internal forces to determine the shape. Using the two scalings derived above in the stiff and floppy regime, we note that $U_f/U_s\sim\kappa^2 k^6/\mu^2 \omega^2=(k\ell)^6=1/a^6$.

\subsubsection{Asymptotics}

Expanding the distributed moment in powers of the small parameter, namely $f=\epsilon f^{(1)}+\epsilon^2f^{(2)}+...$, and furthermore expanding in the same basis as the height field such that $f^{(1)}=f_0\Re\{\sum{f_ne^{in(kx-\omega t)}}\}$, we utilize the results for the pressure and streamfunction from the previous section to find the linear response for the height field as a function of the internal tangential stress
\begin{gather}\label{bn_tang}
b_n=\frac{f_0 n}{\kappa k^2\left[n^3+i2a^3\right]}f_n.
\end{gather}

Using the result Eq.~\eqref{bn_tang}, we are then able to derive the pumping flow,  Eq.~\eqref{velocity}, as a function of the activity, elasticity, and viscosity, and we obtain
\begin{gather}
\label{tangential_velocity}
U^{(2)}=\frac{1}{2}\sum_n{\omega k{|nb_n|^2}}=\frac{1}{2}\frac{\omega f_0^2}{\kappa^2k^3}\sum_n{{\frac{n^4|f_n|^2}{n^6+4a^6}}}\cdot
\end{gather}
In the stiff limit, $a\ll 1$,  the asymptotic results in Eq.~\eqref{tangential_velocity} recover the scaling derived in Sec.~\eqref{sec:scaling_tang}.  For the floppy limit, $a\gg 1$,  the series in Eq.~(\ref{tangential_velocity}) is only asymptotically convergent, but for a finite sum the scaling in Sec.~\eqref{sec:scaling_tang} also holds.

\subsection{Active normal stresses}
\subsubsection{Setup}

In the section above we neglected the details of the activity within the membrane, in favor of a more generic modeling approach describing the relationship between fluid flow,  internally applied bending moments, and passive bending resistance. In a biological context, many sources of activity could instead generate normal stresses in the membrane.  Our second model, described below, considers a concentration of active elements dispersed throughout the membrane and generating fluid stresses.

\begin{figure}[t!]
\includegraphics[width=.6\textwidth]{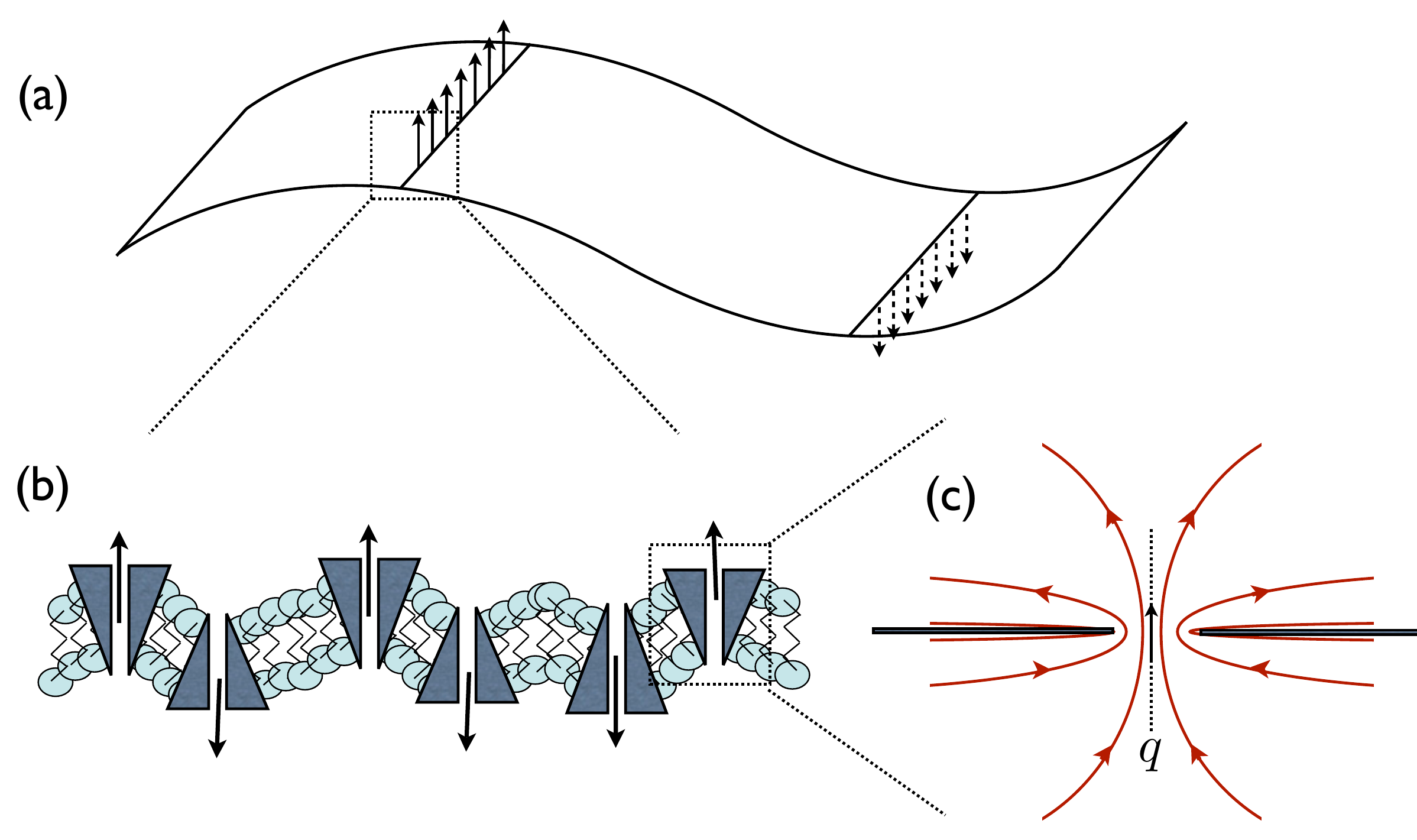}
\caption{\label{schematic_inclusion} Schematic illustration of membrane deformation by active inclusions: (a) Active inclusions embedded in the surface; the inclusions induce flow fields which lead to pressure drop and thus  normal stresses acting on the membrane; (b) Zoomed-in version of the membrane where the size of each inclusion and the local bending of the membrane  are schematically represented; (c) Sketch of the streamlines for a single circular aperture in a flat surface pumping fluid with flow rate $q$;   at leading order the molecular length scale, $d$, is much smaller than the typical membrane scale, $L$, and thus the flow is assumed to be unaffected by membrane curvature.}
\end{figure}

A schematic of the proposed model system is sketched in Fig.~\ref{schematic_inclusion}.  A dilute concentration of ``pumps", each one capable of driving a microscopic flow through the membrane surface, act as inclusions, effectively modifying the material properties.  Not only does the shape of the individual pump alter the shape of the membrane \cite{Ramaswamy:2000p165,Huang:2006p115}, but the flow itself generates fluid stresses on the surface. 

Each pump is modeled as a circular aperture of radius $d$. 
Since $d$ is a molecular length scale far smaller than any other length scale, $L$,  in the system, we can approximate the flow as resulting from a point source embedded in on a flat surface \cite{HappelBrenner}, such that the stream function is given by $\psi=-q/2\pi[1-(\b{\hat{t}}\cdot\b{r}/r)^3]$, where $\b{\hat{t}}$ is the radial tangent vector on the surface, $\b{r}$ is the position of interest in the fluid, and $q$ is the volumetric flow rate through the inclusion.  
The corresponding pressure drop across the aperture is $\delta p=3q\mu/d^3$.  


In order to satisfy the equations of force balance we need to calculate the normal and tangential stress due to not just one pump, but a concentration of inclusions.  Each pump has a preferred direction, and thus we must generally consider the concentration difference, $n=n^{+}-n^{-}$, where $n^{+}$ and $n^{-}$ are the concentrations of pumps pointing in the positive and negative $z$ directions, respectively.  For convenience we will consider the dimensionless quantity $\phi=n/n_0$, where $n_0$ is the equilibrium concentration difference \cite{Ramaswamy:2000p165}.  
 
The normal stress on the membrane due to a single inclusion is simply the pressure drop from the fluid, while the tangential stress on the surface of the membrane decays like $1/\rho^2$, where $\rho=\sqrt{x^2+y^2}$.  The length scale $d$ dominates this contribution, and locally this implies that the tangential stress per length is of the same order as the pressure drop, i.e.~$\b{\hat{t}}\cdot\boldsymbol{\sigma}\cdot\b{\hat{n}}\sim q\mu/d^3$.  However, because the stream function is axisymmetric, the tangential component of the fluid stress integrates to zero over the entire membrane, and thus does not enter the force balance equations. 
 
A general functional describing the enthalpy of the membrane including active pumps is given by
\begin{gather}
G=\int{\frac{\kappa}{2}(C-H_0\phi)^2}dS,
\end{gather}
where $H_0$ a signed measure of the intrinsic curvature for the active elements, and we have neglected effects from 2D compressibility in the concentration, as well as higher order effects coming from gradients in the concentration field \cite{Reigada:2005p75,Ramaswamy:2000p165,Veksler:2007p136}.  

Performing the functional extremization and linearization for the active pump enthalpy, and including the fluid stresses from pump activity, we now find the dynamic equations to be
\begin{subeqnarray}
\label{active_EOM}
	2\kappa\frac{\partial^4h^{(1)}}{\partial x^4}-\kappa H_0\frac{\partial^2 \phi}{\partial x^2}&=&-p^{(1)}-\frac{q\mu}{d^3}\phi  \nonumber\\
	&& -2\mu\left(\frac{\partial^2\psi^{(1)}}{\partial z\partial x}\right)_S,\\
	\tau^{(1)}&=&\mu\left[\frac{\partial^2 \psi^{(1)}}{\partial z^2}-\frac{\partial^2\psi^{(1)}}{\partial x^2}\right]_S. \quad
\end{subeqnarray}
As in the case addressed in the previous section, the tangential stress balance yields zero tension at leading order. 

\subsubsection{Scalings}
\label{scalings_n}

Here again we use scaling arguments to derive the expect form for the macroscopic flow pumped by the membrane.  In addition to the stiff (s) versus floppy (f) regimes explained above, we  must consider in addition the competition between by the spontaneous curvature and the deformation induced by the active pumping mechanism:  In one limit the local stiffness introduced by the molecular curvature of the inclusions overrides the pumping activity (we denote this limit h), while in the opposite limit the spontaneous curvature is dominated by the pump activity (denoted a). We have thus four different limits to characterize. 

Let us denote by $\phi_0$ the typical magnitude of the dimensionless concentration of pumps, and the typical force generated by the pumps as $f_{act}=q\mu/d$. To measure the competition between the natural curvature of the inclusions (h case) and the one arising from the  activity-induced fluid flow (a case), we introduce the  dimensionless parameter, $A=H_0\kappa d^2k^2/f_{act}$.

For stiff membranes ($a \ll  1$), in the limit where the bending from activity is predominant, i.e.~$A\ll 1$, force balance reveals that $b_{{\rm eff}}\sim f_{act}\phi_0/\kappa d^2k^4$,  while in the opposite limit where the bending arises from molecular curvature ($A \gg  1$), we get  $b_{{\rm eff}}\sim \phi_0H_0/k^2$. In contrast, for floppy membranes ($a \gg  1$), the case of active inclusions ($A \ll  1$) leads to the scaling $b_{{\rm eff}}\sim f_{act}\phi_0/\mu\omega_0d^2k$, while in the limit where the inclusions pump a very small amount of fluid transverse to the membrane ($A\gg 1$), we obtain $b_{{\rm eff}}\sim H_0\phi_0\kappa k/\mu\omega$. 

Now, the expected fluid  velocities in the four different limits can be found by again using the analogy with the swimming sheet, $U \sim c (b_{{\rm eff}}k)^2$. For stiff active membranes ($a \ll  1$, $A \ll 1$), we expect $U_{sa} \sim \omega (f_{act}\phi_0)^2/\kappa^2d^4k^7$, while stiff inactive membranes ($a \ll  1$, $A \gg 1$) should lead to $U_{sh} \sim \omega (H_0\phi_0)^2/k^3$. In the inactive case we note that the fluid velocity no longer depends on the membrane stiffness, as the intrinsic curvature $H_0$ governs the bending penalty at the same order in $\kappa$ as local deformations in the height field.

In the case of floppy active membranes ($a \gg  1$, $A \ll 1$), we expect to obtain $U_{fa}\sim (f_{act}\phi_0)^2/\mu^2d^4\omega k$, while for inactive floppy membranes ($a \gg  1$, $A \gg 1$)  the pumping flow should scale like $U_{fh} \sim (H_0\phi_0\kappa)^2k^3/\mu^2\omega$.  It is notable that even in the inactive case, the mismatch of curvature between the inclusions and the elastic membrane they are embedded in can, alone, lead to deformation that gives rise to fluid transport; even in the floppy limit consequences of the bending rigidity $\kappa$ cannot be neglected.

\subsubsection{Asymptotics}

Using the Fourier decomposition for the concentration of inclusions,  $\phi(x,t)=\sum{\phi_ne^{in(kx-\omega t)}}$, the linear response of 
Eq.~\eqref{active_EOM}  is found to give
\begin{gather}
n^4k^4 b_n+\frac{i2\mu\omega nk}{\kappa}b_n=-H_0n^2k^2\phi_n-\frac{f_{act}}{\kappa d^2}\phi_n.
\end{gather}

The final linear response for the height takes the form
\begin{gather}\label{lr_normal}
b_n=-\frac{f_{act}}{\kappa k^4 d^2}\frac{1+An^2}{n^4+i2a^3n}\phi_n.
\end{gather}

Plugging Eq.~\eqref{lr_normal} into  Eq.~(\ref{velocity}) we finally find that the macroscopic velocity is given by
\begin{gather}\label{final_n}
U^{(2)}=\frac{1}{2}\sum_n{\omega k |nb_n|^2}=\sum_n\frac{1}{8}\frac{\omega f_{act}^2}{k^7d^4\kappa^2}\left[\frac{(1+An^2)^2}{n^8+4n^2a^6}\right]|\phi_n|^2.
\end{gather}
In the stiff ($a \ll 1$) and floppy ($a\gg 1$) limits, as well as the limits where intrinsic pump curvature dominates ($A\gg 1$) or is dominated by ($A\ll 1$) deformation from the active normal stresses, the final asymptotic  results in Eq.~\eqref{final_n} confirm all the scaling predictions in Sec.~\ref{scalings_n}. 

\section{Discussion}

In summary, although the framework for characterizing fluid transport and locomotion by a waving sheet has existed since the 50's, in this  work we have  attempted to go beyond a prescription of surface deformation by instead prescribing internal activity (so starting from dynamics instead of kinematics). Both membrane deformation and fluid transport can then be solved by solving a dynamic balance between activity, passive resistance, and external fluid stresses. 
We have used two models to cover a range of possible forcing, namely a planar distribution of bending moments  that generate normal deformation, and a simple model of active constituents that produce normal permeative flow, resulting in sheet undulation.  

From an experimental standpoint, what is the typical magnitude of the flow which could be induced by active  mechanisms  similar to the ones described in this paper? For lipid bilayers, bending rigidities are on the order of $\kappa\sim 10^{-19} Nm$ \cite{ZimmerbergKozlov2006}, and using cross-linked molecular motors as one model microscopic force generator, a single molecular machine could generate forces on the order of $\sim 1pN$ \cite{Finer1994}.  If these were distributed throughout a membrane, say with a dimensionless concentration of $\phi\sim 10^{-3}$, we could expect a magnitude for the internal moment per unit length of $f_0\sim 10^{-15} N$. On cellular length scales $L\sim 100\mu m$, with $k\sim 1/L$, the range of frequencies $\omega\sim 10^0-10^2 Hz$  could include both the  stiff and floppy regimes, and as a result we could expect macroscopic velocities  on the order of $U\sim 1\mu m/s$ for low frequencies (stiff limit) or $U\sim 1-100 \mu m/s$ for higher frequencies (floppy regime).

For transmembrane proteins capable of inducing a microscopic flow through a surface, such as aquaporins or proton pumps, the volumetric flow rate is difficult to estimate, but we can use previous simulation results for guidance \cite{Zhu2004,Davis2007}.  For membrane constituents such as lipids or proteins a typical radius of gyration gives $H_0\sim 1 nm^{-1}$ \cite{ZimmerbergKozlov2006}.  This yields a value for the parameter $A\sim (10^{-18}N)/f_0$. 
For  molecular motors generating fluid flow normal to the membrane with a force per motor on the order of $f_0\sim 1pN$, this makes $A\ll 1$, i.e.~the active limit; for aquaporins or other active pores that are not designed specifically to move cellular structures, $A\gg 1$.  With a frequency of oscillation of $\omega\sim 1 Hz$, these membranes are in the stiff limit. With a dimensionless concentration as small as $\phi_0\sim 10^{-3}$, the macroscopic pumping velocity can be as large as $U\sim 10-100 \mu m/s$ for the active case, and $U\sim 1\mu m/s$ for inactive membranes.

One possible experimental realization for a self-propelled active membrane could be in the form of a closed bilayer vesicle with embedded active pumps.  For a spherical vesicle of radius $R$ and  wavelength undulations satisfying $\lambda \gg R$, we can use the above calculations in tandem with the swimming results of Stone and Samuel \cite{ss96} to get an estimate of the vesicle swimming speed
\begin{gather}
\label{SamuelStone}
U\b{\hat{z}}\approx -\frac{1}{4\pi R^2}\int_S{\b{u}dS},
\end{gather}
where $\b u=U^{(2)}\b t$ is the local fluid velocity created by the activity-induced membrane deformation; up to a geometric constant, we thus get that  the instantaneous swimming velocity of this active vesicle is the same as that given in our calculations above.  Several previous studies have examined the possibility of self-propelled vesicles \cite{MiuraVesicleMotion2010,Evans2010,howse07}, and our results connecting the internal stress state to macroscopic motion can thus be used as a probe of the activity. One  could envision a situation where the diffusivity of active vesicles would be experiementally measured; in the presence of active pumps, this diffusivity would  be enhanced by the propulsion velocity as $D_{eff}\sim U^2/D_r$, where $D_r$ is the vesicle rotational diffusion \cite{Berg93}, which could then  be directly related  to the activity via the results derived in this paper.  Our framework could serve, for example,  as a way to rule out specific forms of activity in a membrane.

\section*{Acknowledgements}
This work was supported in part by the National Science Foundation (Grant CBET-0746285 to E.L.).

\bibliography{active_membrane}

\end{document}